\documentclass{article}
\usepackage{booktabs}
\usepackage{multirow}
\usepackage{threeparttable}
\usepackage{microtype}
\usepackage{graphicx}
\usepackage{subcaption}
\usepackage{booktabs} 
\usepackage[table]{xcolor}
\usepackage{hyperref}

\usepackage[preprint]{icml2026}

\usepackage{amsmath}
\usepackage{amssymb}
\usepackage{mathtools}
\usepackage{amsthm}
\usepackage{longtable}
\usepackage{booktabs} 
\usepackage{multirow}

\usepackage[capitalize,noabbrev]{cleveref}

\theoremstyle{plain}

\theoremstyle{definition}

\theoremstyle{remark}

\usepackage[textsize=tiny]{todonotes}

\icmltitlerunning{Submission and Formatting Instructions for ICML 2026}

\begin{document}

\twocolumn[
  \icmltitle{CRCC: Contrast-Based Robust Cross-Subject and Cross-Site\\
  Representation Learning for EEG \\}



\icmlsetsymbol{equal}{*}
\icmlsetsymbol{corresponding}{+}

\begin{icmlauthorlist}
\icmlauthor{Xiaobin Wong}{aff1,equal}
\icmlauthor{Zhonghua Zhao}{aff3,aff4,equal}
\icmlauthor{Haoran Guo}{aff5,equal}
\icmlauthor{Zhengyi Liu}{aff6,equal}

\icmlauthor{Yu Wu}{aff7}
\icmlauthor{Feng Yan}{aff8,aff9}
\icmlauthor{Zhiren Wang}{aff8,aff9}
\icmlauthor{Sen Song}{aff1,aff2}

  \end{icmlauthorlist}

\icmlaffiliation{aff1}{Tsinghua Laboratory of Brain and Intelligence}
\icmlaffiliation{aff2}{School of Biomedical Engineering, Tsinghua University}
\icmlaffiliation{aff3}{Institute of Automation, Chinese Academy of Sciences}
\icmlaffiliation{aff4}{University of Chinese Academy of Sciences}
\icmlaffiliation{aff5}{Weixian College, Tsinghua University}
\icmlaffiliation{aff6}{School of Artificial Intelligence, Beijing University of Posts and Telecommunications}
\icmlaffiliation{aff7}{School of Computer Science and Technology, Northwestern Polytechnical University, Xi'an, China}
\icmlaffiliation{aff8}{Beijing Huilongguan Hospital, Capital Medical University}
\icmlaffiliation{aff9}{Peking University Huilongguan Clinical Medical School}

\icmlcorrespondingauthor{Zhiren Wang, Sen Song}{sensong@tsinghua.edu.cn}

  \vskip 0.3in
]



\printAffiliationsAndNotice{}  

\begin{abstract}
EEG-based neural decoding models often fail to generalize across acquisition sites due to structured, site-dependent biases implicitly exploited during training. We reformulate cross-site clinical EEG learning as a bias-factorized generalization problem, in which domain shifts arise from multiple interacting sources. We identify three fundamental bias factors and propose a general training framework that mitigates their influence through data standardization and representation-level constraints. We construct a standardized multi-site EEG benchmark for Major Depressive Disorder and introduce CRCC, a two-stage training paradigm combining encoder–decoder pretraining with joint fine-tuning via cross-subject/site contrastive learning and site-adversarial optimization.CRCC consistently outperforms state-of-the-art baselines and achieves a 10.7 percentage-point improvement in balanced accuracy under strict zero-shot site transfer, demonstrating robust generalization to unseen environments.
\end{abstract}

\section{Introduction}
Human experiences are inherently heterogeneous, and such individual differences constitute a major source of variability in neural signals. However, this variability poses significant challenges for neural data analysis: algorithms must contend not only with structural brain differences, but also with diverse functional expression patterns across individuals. Consequently, many traditional algorithms exhibit a substantial degradation in performance when evaluated on cross-subject data. To address this issue, several recent studies have proposed methods for extracting cross-subject neural representations \cite{shen2022contrastive, zhao2021plug}, partially alleviating the impact of individual variability

However, beyond individual differences, the site effect remains another formidable “dark cloud,” obscuring the clinical translation of AI-driven biomarkers from bench to bedside. The vast majority of existing studies still rely on leave-one-site-out (LOSO) cross-validation or few-shot learning paradigms \cite{wu2020electroencephalographic, an2023dual}, while only a handful of works have attempted to evaluate model generalization under cross-dataset zero-shot protocols \cite{zhang2025multi}.
This discrepancy strongly suggests that many models may capture spurious “shortcuts” for label discrimination rather than genuine neural signatures \cite{zhao2025predicting}. In particular, when healthy controls and patients are recruited from different environments, inflated accuracy estimates are often confounded by noise introduced by heterogeneous acquisition conditions.

Relatedly, cross-dataset passive brain–computer interface (pBCI) decoding studies that integrate heterogeneous paradigms—such as affective computing—inevitably suffer from discrepancies in stimulus materials across datasets when evaluated in zero-shot settings \cite{zheng2015investigating, chen2023large}.
In particular, during cross-paradigm generalization, performance gains measured by clustering-based metrics must be carefully interpreted, as they may arise from differences in low-level perceptual responses (e.g., visual cortex activation) induced by the stimuli themselves rather than task-relevant neural representations. Consequently, when investigating site effects, adopting simplified paradigms such as resting-state recordings is of particular importance.

\begin{figure*}[t] 
    \centering
    \includegraphics[width=0.95\textwidth]{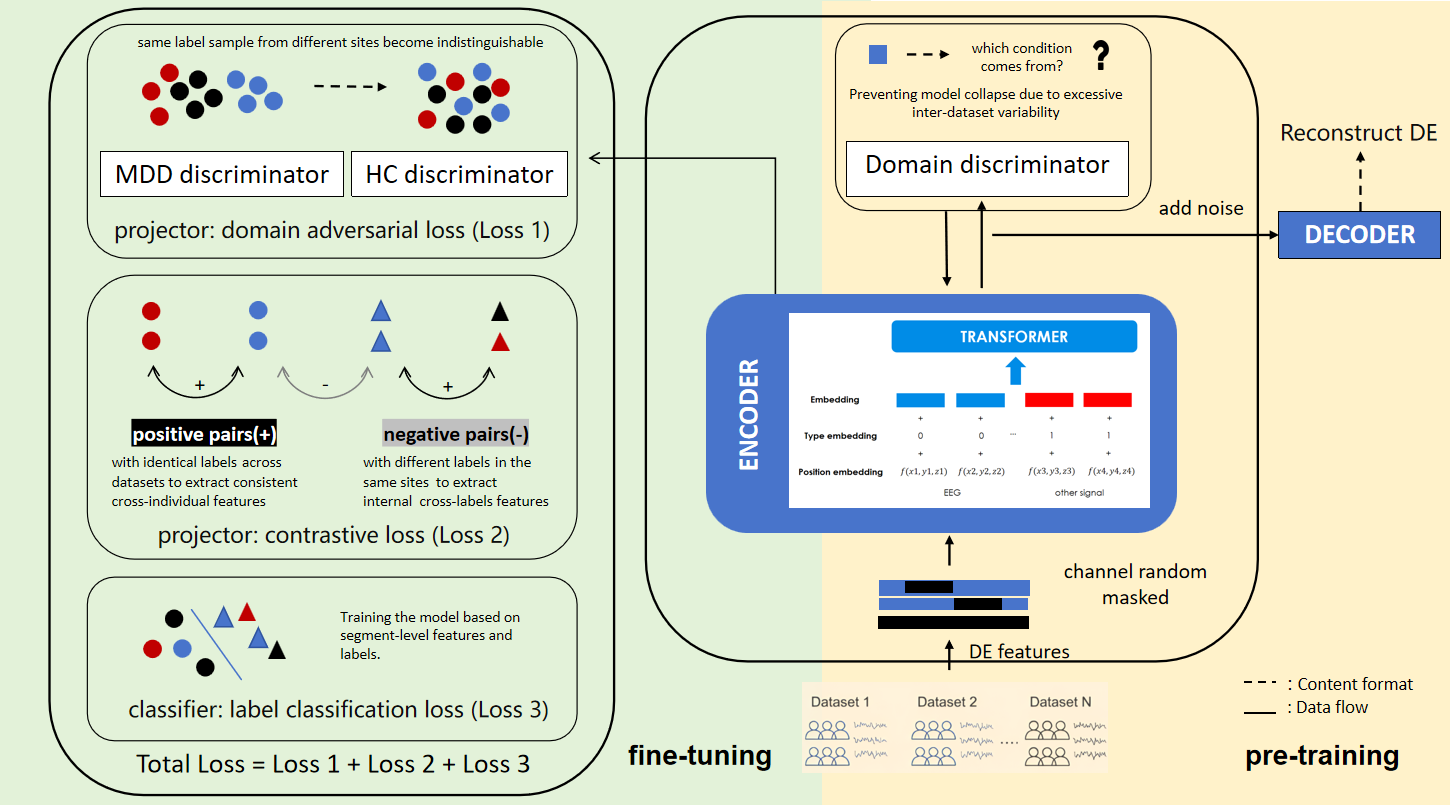} 
    \caption{The CRCC Framework: our model comprises two primary stages: the pre-training phase (right) and the fine-tuning phase (left). During pre-training, the model takes 10-second EEG Differential Entropy (DE) features as input. These features are processed by an encoder, with gradients backpropagated through a joint objective involving a noise-reducing decoder and a domain discriminator, the latter of which facilitates robust representation learning across multiple open-source datasets. In the subsequent fine-tuning stage, the decoder and domain discriminator are detached. The core fine-tuning objective integrates a standard cross-entropy loss for classification with two specialized components: a contrastive learning module designed for cross-subject feature extraction and an adversarial loss aimed at mitigating inter-site variability.}
    \label{fig:main_architecture}
\end{figure*}

To effectively address the challenges of cross-site generalization, we hypothesize that the observed total distributional shift, denoted as $\mathcal{B}_{total}$, is a composite effect of multi-level confounding factors. Formally, we decompose the domain bias into three primary components:
\begin{equation}
\mathcal{B}_{total} = \mathcal{B}_{sys} + \mathcal{B}_{pop} + \mathcal{B}_{diag} \label{eq:total_errors}
\end{equation}
where:
\begin{itemize}
    \item $\mathcal{B}_{sys}$ represents \textbf{systematic operational bias}, arising from site-specific hardware configurations, electrode impedances, and environmental noise during recording.
    \item $\mathcal{B}_{pop}$ denotes \textbf{population-level variance}, reflecting the demographic and biological heterogeneity across different geographical recruitment sites.
    \item $\mathcal{B}_{diag}$ signifies \textbf{diagnostic uncertainty}, referring to the inherent variance in clinical labeling and the overlapping neural signatures between different mental health states.
\end{itemize}

For any adjustment made to address inter-site differences, we explicitly map it to this formulation and aim to minimize the corresponding bias term. Under this formulation, the core objective of our framework is to systematically decouple and mitigate each component.

In this study, we first constructed a disease classification dataset (consisting of patients with major depressive disorder (MDD) and healthy controls (HC)) across seven sites using the same acquisition equipment. This allows us to systematically minimize the first and last error terms in Eq. (1).Then, based on this multi-site dataset, we designed an algorithmic framework (shown in (Fig.\ref{fig:main_architecture})) to mitigate both individual differences and site effects, which corresponds to the middle error term in Eq. . This framework demonstrates the feasibility of cross-site disease classification, offering potential for future population-oriented disease screening based on neuro-biomarkers initiated by various researchers.

Our main contributions are as follows.\\
\textbf{By formalizing the error, we enable its systematic reduction through targeted external interventions.} In other words, we constrain how methods should be designed by reducing bias sources. The error formulation further guides the construction of our data cohorts and the algorithm design. 

\textbf{Multi-dataset pre-training}: We designed a masked autoencoder (MAE) and decoder approach to extract neural data features during the pre-training stage. 

\textbf{Cross-subject and cross-site alignment loss:} Drawing inspiration from the previous studies, we implemented a contrastive learning based strategy that extract subject-invariant features of the same label across different sites. 

\textbf{Sites-adversarial loss}: We performed site-based adversarial training for each label separately, requiring the model to be unable to distinguish which site a specific sample originates from.

We focus on the critical generalization challenges that severely hinder the clinical translation of EEG technologies. By employing a one by one controlled-variable approach to systematically eliminate noise components, we aim to provide valuable insights and inspirations for future framework development in this field.

\section{Related Work}
\subsection{EEG dataset for site effects exploration: where the bias comes from?}

Deep learning algorithms and datasets for EEG decoding are no longer rare, with large-scale databases such as PREDICT\cite{cavanagh2019multiple} and TDBRAIN\cite{van2022two} becoming increasingly available.\\However, for a long time, algorithms have largely failed to achieve true clinical utility. A primary reason is that most models fail to demonstrate robust generalization in real-world scenarios. We remain uncertain whether this bias stems from the noise inherent in the labels themselves—previous research indicates that diagnostic consistency, for example, MDD is only 0.28\cite{regier2013dsm}—or from systematic operational errors. To effectively control the noise at these sites, we have to constructed a new multi-site dataset.\\
Compared to EMBARC\cite{trivedi2016establishing} and CANBIND\cite{lam2016discovering}, the majority of our data was collected using identical equipment. The Assessments at Sites 3, 4, 5, and 6 were conducted by a consistent team of evaluators to ensure diagnostic uniformity. Furthermore, to avoid the common fallacy in previous studies in which healthy controls (HC) and patients were collected in separate batches or settings, we interleaved the recruitment and data collection of HCs and MDD patients in site 5 under identical conditions throughout the process. \\In contrast, data from Site 1 were collected by collaborators following traditional acquisition practices in batches: specifically, collecting patient data in a concentrated period followed by the centralized recruitment of healthy controls. 
\textbf{This practice directly exacerbates the risk of amplified} \textbf{$\mathcal{B}_{sys}$ and $\mathcal{B}_{pop}$ effects}. We intentionally retained this cohort to validate our hypothesis regarding these artifacts.\\
Information for our dataset is presented in Table \ref{table:demographics_final}, and a detailed comparison with existing databases is provided in Table \ref{table:dataset_comparison}. 

Overall, our database provides the necessary foundation for the development of more robust clinical algorithms.

\begin{table*}[t]
\centering
\begin{threeparttable} 
\caption{Demographic Information of the Multi-site Dataset.}
\label{table:demographics_final}
\vskip 0.05in
\begin{small}
\begin{sc}
\begin{tabular}{lcccccccr}
\toprule
\multirow{2}{*}{Site Code} & \multicolumn{3}{c}{HC} & \multicolumn{3}{c}{MDD} & \multirow{2}{*}{Device} \\
\cmidrule(r){2-4} \cmidrule(l){5-7}
 & N & BDI & HAMD & N & BDI & HAMD &  \\
\midrule
0  & 148 & 8.53$\pm$7.57 & / & 0 & / & / & \multirow{7}{*}{\begin{tabular}[c]{@{}c@{}}32-channel\\ EEG device\end{tabular}} \\
1\tnote{*} & 40  & / & / & 92 & / & 24.37$\pm$5.33 &  \\
2  & 186 & / & / & 0  & / & / &  \\
3  & 2   & / & / & 70 & 24.45$\pm$12.96 & 19.32$\pm$7.37 &  \\
4  & 3   & 5.33$\pm$5.50 & 3.33$\pm$3.06 & 216 & 18.08$\pm$13.09 & 17.22$\pm$8.74 &  \\
5  & 32  & 4.63$\pm$5.68 & 4.13$\pm$5.42 & 28  & 16.26$\pm$11.68 & 14.00$\pm$8.53 &  \\
6  & 0   & / & / & 10  & 26.73$\pm$12.79 & 19.87$\pm$9.80 &  \\
\midrule
All & 411 & - & - & 416 & - & - &  \\
\bottomrule
\end{tabular}
\end{sc}
\end{small}

\begin{tablenotes} 
    \footnotesize
    \item[]\hspace*{-15pt}\textbf{1*}: Demographic statistics for the dataset are currently in progress; partially completed scale entries are presented.
    \item[]\hspace*{-15pt}'/': Indicates clinical scale or data was not collected at this site.
    \item[] \hspace*{-15pt}\textbf{BDI}: Beck Depression Inventory (scores 0--13: no depressive symptoms). Values: Mean $\pm$ SD.
    \item[]\hspace*{-15pt}\textbf{HAMD/HDRS}: Hamilton Depression Rating Scale (17-item, scores 0--7: no depressive symptoms). 
\item[]\hspace*{-10pt}

\end{tablenotes} 
\end{threeparttable} 
\end{table*}

\begin{table*}[t] 
\centering
\caption{Comparison Between Our Dataset and Existing EEG Databases for MDD.}
\label{table:dataset_comparison}
\vskip 0.05in
\begin{small}
\begin{sc}
\begin{tabular}{lccccc}
\toprule
Dataset & MDD (N) & HC (N) & Multi-site & Same Device & Same Evaluators \\
\midrule
PREDICT & 46 & 75 & $\times$ & $\checkmark$ & $\checkmark$ \\
Tdbrain & 426 & 47 & $\times$ & $\checkmark$ & $\checkmark$ \\
EMBARC  & 309 & 40 & $\checkmark$ & $\times$ & $\times$ \\
CANBIND & 211 & 112 & $\checkmark$ & $\times$ & $\times$ \\
\midrule
\textbf{Ours} & \textbf{411} & \textbf{416} & \textbf{$\checkmark$} & \textbf{$\checkmark$} & \textbf{$\checkmark$} \\
\bottomrule
\end{tabular}
\end{sc}
\end{small}
\end{table*}


\subsection{Rethinking Foundation Models for EEG}
The landscape of EEG decoding has been recently reshaped by the emergence of foundation models, such as LaBraM\cite{jiang2024large} and EEGPT\cite{wang2024eegpt}, which leverage massive datasets and self-supervised learning to capture intricate neural representations. These models primarily follow a scaling-law trajectory, where escalating model complexity is prioritized to exhaustively learn high-dimensional features. However, we argue that in real-world clinical applications, performance is often bottlenecked not by a lack of model capacity, but by the intrinsic domain shifts—specifically, the substantial inter-subject variability and site-specific effects inherent in neural recordings.\\
The prevalent failure of these high-capacity models in zero-shot generalization suggests a critical vulnerability: during the convergence of complex loss functions, models may inadvertently assimilate non-neural artifacts and site-specific noise that are spuriously correlated with labels. Unlike the prevailing trend of increasing parameter scale, our work advocates for model parsimony.

\section{Multi-dataset pretraining}
\subsection{Motivation}
\textbf{Core Objective of Pre-training.} Our pre-training phase is designed to achieve one pivotal goal: to endow the model with the intrinsic capability to discern salient neural information from pervasive noise. Building upon the standard MAE framework, we strategically incorporate a specialized decoder and domain-adversarial training to suppress inherent noise. 

\subsection{Pre-training with Masked Reconstruction}
The pre-training strategy employs MAE framework to learn robust representations from multi-channel neural signals. The model processes combined EEG data through a unified transformer architecture, where input features are organized as $\mathbf{F} \in \mathbb{R}^{B \times C \times n_{\text{second}} \times d_{\text{feature}}}$ for a batch with $C$ channels. To enable the model to distinguish between modalities and encode spatial information, we incorporate three complementary embeddings. A type embedding differentiates EEG and other signal channels through a binary indicator $\mathbf{T} \in {0,1}^{B \times C}$. Spatial position embeddings encode the 3D coordinates $\mathbf{L} \in \mathbb{R}^{B \times C \times 3}$ of each electrode in standardized space using sinusoidal functions applied independently to each dimension.

\subsection{EEG Decoder}
For reconstruction, the encoder representations are decoded back to the original feature space through modality-specific output heads, ensuring that EEG signals are reconstructed with appropriate transformations. The reconstruction loss is computed exclusively on masked and valid positions using smooth L1 loss, which provides robustness to outliers in neural recordings. Specifically, for the set of masked and non-padded positions $\Omega = {(i,j,t) : \mathbf{M}{i,j,t} = 1 \wedge \mathbf{P}{i,j} = 0}$, the reconstruction loss is defined as
$$\mathcal{L}{\text{rec}} = \frac{1}{|\Omega|} \sum{(i,j,t) \in \Omega} \rho_\beta(\hat{\mathbf{F}}{i,j,t} - \mathbf{F}{i,j,t}),$$
where $\mathbf{M}$ denotes the mask indicator, $\mathbf{P}$ denotes the padding mask, and $\rho_\beta$ is the smooth L1 function with parameter $\beta$. 

\subsection{Domain discriminators in Pre-training}
To address the domain shift challenge arising from diverse acquisition protocols, equipment and even paradigms across datasets, we employ domain adversarial training. The aggregate representation $\mathbf{z} \in \mathbb{R}^{B \times d_{\text{embed}}}$ obtained by pooling across non-padded channel tokens is processed through a gradient reversal layer $\text{GRL}\alpha$ before being fed to multiple domain discriminators ${D_k}{k=1}^{K}$, each targeting a specific aspect of domain variability such as data source, electrode density, or recording equipment. The adversarial loss is formulated as
$$\mathcal{L}{\text{adv}} = \sum{k=1}^{K} \mathbb{E}{(\mathbf{z}, y_k) \sim \mathcal{D}} \left[ \mathcal{L}{\text{CE}}(D_k(\text{GRL}_\alpha(\mathbf{z})), y_k) \right],$$
where $\mathcal{L}_{\text{CE}}$ denotes cross-entropy loss with label smoothing and $y_k$ represents the domain label for the $k$-th domain partition. This adversarial mechanism encourages the encoder to learn features that are invariant to domain-specific characteristics while remaining discriminative for the reconstruction task.The overall pre-training objective integrates reconstruction fidelity with domain invariance as
$$\mathcal{L} = \mathcal{L}{\text{rec}} + \lambda{\text{adv}} \mathcal{L}_{\text{adv}}.$$

\section{Targeting Robust Neural Representations via Fine-tuning}
\label{sec:finetuning}

\subsection{Motivation}
\textbf{Fine-tuning Philosophy.} The essence of our fine-tuning strategy lies in a fundamental shift from knowledge acquisition to feature alignment. Having broad neural representations during the pre-training phase, the objective is no longer to pursue the maximal complexity of features. Instead, we aim to guide the model toward distilling domain-invariant signatures by decreasing the errors in \ref{eq:total_errors} . 

\subsection{Contrast-based Cross-subject and Cross-site Loss}
\textbf{This is aimed to minimize the population errors.}\\
The objective of our contrastive loss is to distill neurobiological representations that remain invariant across individuals and recording environments. Let the feature representation extracted by the backbone network for a sample $i$ be $\mathbf{h}_i = f(x_i) \in \mathbb{R}^{F}$. The projected embedding used for contrastive learning is defined as $\mathbf{z}_i = \text{Projector}(\mathbf{h}_i)$.

The philosophy is to minimize the subject-specific noises and site-specific bias. Specifically, selecting positive samples from the same site or negative samples across different sites introduces spurious correlations—namely, an artificial consistency bias between positive pairs and a discrepancy bias between negative pairs. To mitigate this, we propose a \textbf{site-agnostic sampling strategy}: we prioritize cross-site individuals when selecting positive pairs and within-site individuals when selecting negative pairs. 

For an anchor sample $\mathbf{z}_i^a$, let $\mathbf{z}_i^+$ be its positive counterpart (from the same class but a different site) and $\{\mathbf{z}_j^-\}_{j=1}^{M}$ be a set of negative samples from different classes. Using the negative $L_2$ distance as the similarity metric, the contrastive loss for a batch of $N$ samples is formulated as:
\begin{equation}
\mathcal{L}_{\text{con}} = -\frac{1}{N}\sum_{i=1}^{N} \log \left[ \frac{e^{-\|\mathbf{z}_i^{a} - \mathbf{z}_i^{+}\|_2}}{e^{-\|\mathbf{z}_i^{a} - \mathbf{z}_i^{+}\|_2} + \sum_{j=1}^{M} \mathbb{I}_{ij} \cdot e^{-\|\mathbf{z}_i^{a} - \mathbf{z}_j^{-}\|_2}} \right]
\end{equation}

where $\|\cdot\|_2$ denotes the $L_2$ norm. The indicator function $\mathbb{I}_{ij} \in \{0, 1\}$ equals $0$ if the negative sample $\mathbf{z}_j^-$ shares any clinical label with the anchor $\mathbf{z}_i^a$, thereby masking out trivial negatives and focusing the model on authentic category-level discrepancies.

\subsection{Site-adversarial Loss}
This is aimed to decrease all three factors at the same time.\\
The underlying rationale for the site-adversarial loss is that within a unified cultural context, the clinical manifestations of MDD should remain consistent across different sites. Consequently, significant feature disparities across sites are likely a reflection of site-specific noise rather than neural pathology. To counteract this, we employ a \textbf{Site-specific Domain Adversarial} approach.

To distill site-invariant representations, we utilize a Gradient Reversal Layer (GRL) coupled with domain discriminators. The GRL acts as an identity transform during the forward pass but scales the gradient by $-\alpha$ during backpropagation. To account for the distinct distribution shifts within clinical groups, we deploy separate discriminators for the HC and MDD populations, denoted as $D_{HC}$ and $D_{MDD}$, respectively. For a feature $\mathbf{h}_i$, the adversarial objective is to minimize the site-prediction success, formulated as:

\begin{equation}
\mathcal{L}_{\text{adv}} = -\frac{1}{N} \sum_{i=1}^{N} \sum_{d=1}^{S} s_i^{(d)} \log \left( D_{t}(\text{GRL}(\mathbf{h}_i)) \right)
\end{equation}

where $s_i^{(d)}$ is the one-hot ground-truth site label, $S$ is the total number of sites, and $t \in \{\text{HC}, \text{MDD}\}$ selects the corresponding discriminator based on the subject's clinical category. This adversarial mechanism encourages the backbone to suppress site-specific artifacts.

\subsection{Total Loss}

The primary objective of our framework is robust clinical feature extraction for neuro signal (i.e., HC vs. MDD). A classifier head predicts the label distribution $\hat{y}_i = \text{Softmax}(\text{Linear}(\mathbf{h}_i))$. To address the class imbalance inherent in multi-site clinical data, we adopt a weighted cross-entropy loss:

\begin{equation}
\mathcal{L}_{\text{cls}} = -\frac{1}{N}\sum_{i=1}^{N} \sum_{c=1}^{2} w_c \cdot y_{i,c} \log \hat{y}_{i,c}, \quad w_c = \frac{N}{2 \cdot N_c}
\end{equation}

where $w_c$ is inversely proportional to the sample count $N_c$ of class $c$. The final training objective is a collaborative optimization of the contrastive, adversarial, and classification losses:

\begin{equation}
\mathcal{L}_{\text{total}} = \lambda_1 \mathcal{L}_{\text{con}} + \lambda_2 \mathcal{L}_{\text{adv}} + \mathcal{L}_{\text{cls}}
\end{equation}

where $\lambda_1$ and $\lambda_2$ are hyper-parameters that balance the trade-off between feature discriminability and site invariance.

\subsection{Evaluation Strategy: Subject-level Metrics}

To ensure clinical relevance and a fair comparison across diverse architectures, we standardize the evaluation protocol. While all models are trained using their respective loss functions, the final performance is reported using the \textbf{subject as the minimum unit of analysis}. 

For a given subject $k$, let $\{\mathbf{x}_1, \dots, \mathbf{x}_m\}$ be the set of EEG segments extracted from their recording. The subject-level prediction $\hat{Y}_k$ is determined by aggregating the segment-level probabilities:
\begin{equation}
\hat{Y}_k = \text{argmax} \left( \frac{1}{m} \sum_{j=1}^{m} P(y | \mathbf{x}_j; \theta) \right)
\end{equation}
All balanced accuracy, F1 score, AUC, recall and precision results presented in Sec.~\ref{table:main_results} are calculated based on these aggregated subject-level predictions. This approach mitigates the potential bias introduced by segment-level noise and aligns with the practical requirements of clinical diagnosis.

\section{Experiment}
\label{sec:experiment}
\subsection{Experimental Setup}
\label{sec:setup}

\textbf{Data processing.} All EEG recordings were processed following a standardized pipeline, including downsampling, band-pass filtering, and Independent Component Analysis (ICA) for artifact removal. Details of the paradigm and preprocessing stages are provided in Appendix \ref{appendix:paradigm}. The pre-training phase utilized a collection of entirely open-source datasets, the specifics of which are documented in Appendix \ref{appendix:hyperparams} and \ref{appendix:pretrain}.

\textbf{Dataset Partitioning and Leakage Prevention.} To ensure the integrity of our evaluation, all data partitioning during the fine-tuning stage was conducted on a \textbf{per-subject basis}. Strict protocols were implemented to prevent any potential \textbf{subject leakage}; for instance, multiple sessions from the same individual (e.g., longitudinal follow-ups) were strictly assigned to either the training or validation set as a single block. 

\textbf{Zero-shot Verification.} We designated Sites 1 and 5 as independent test sets for cross-site zero-shot verification, as these sites exhibit a relatively balanced distribution of HC and MDD subjects. During the fine-tuning stage, we performed 10-fold cross-validation on the remaining sites. The model weights yielding the highest average validation accuracy were selected and evaluated on the held-out test sites without further adaptation.

\textbf{Condition Separation.} Recognizing that neural dynamics differ significantly between \textbf{Eyes-Open (EO)} and \textbf{Eyes-Closed (EC)} states, we treated these two conditions as distinct sub-pipelines throughout the fine-tuning and evaluation phases.

\subsection{Results of subject-independent classification}
Rather than merely improving accuracy, our experiments diagnose why existing models fail under cross-site shift. 
\begin{table*}[t]
\centering
\caption{Performance Comparison on Independent Test Sites (Site 1 \& Site 5). "EO" and "EC" denote Eyes-Open and Eyes-Closed states, respectively. B-Acc and ZS-BAcc represent Balanced Accuracy and Zero-shot Balanced Accuracy.}
\label{table:main_results}
\vskip 0.05in
\begin{small}
\begin{sc}
\begin{tabular}{llcccccc}
\toprule
Site \& State & Method & B-Acc & ZS-BAcc & Precision & Recall & F1 Score & AUROC \\
\midrule
\multirow{4}{*}{site1-EO} & DE+MLP & .819 $\pm$ .06 & .570 $\pm$ .01 & \textbf{.815 $\pm$ .02} & .819 $\pm$ .06 & .810 $\pm$ .06 & \textbf{.874 $\pm$ .02} \\
 & LaBraM & .783 $\pm$ .22 & .504 $\pm$ .18 & .594 $\pm$ .41 & .676 $\pm$ .46 & .631 $\pm$ .43 & .791 $\pm$ .25 \\
 & EEGPT &.815 $\pm$ .05 & .491 $\pm$ .01 & .755 $\pm$ .05 & .716 $\pm$ .04 & .709 $\pm$ .04 & .766 $\pm$ .06 \\
 & \textbf{Ours} & \textbf{.832 $\pm$ .06} & \textbf{.645 $\pm$ .03} & .799 $\pm$ .06 & \textbf{.838 $\pm$ .03} & \textbf{.815 $\pm$ .05} & .870 $\pm$ .04 \\
\midrule
\multirow{4}{*}{site1-EC} & DE+MLP & .818 $\pm$ .05 & .537 $\pm$ .01 & .818 $\pm$ .04 & .818 $\pm$ .05 & .814 $\pm$ .05 & .869 $\pm$ .05 \\
 & LaBraM & .831 $\pm$ .18 & .541 $\pm$ .16 & .744 $\pm$ .39 & .696 $\pm$ .37 & .720 $\pm$ .38 & .840 $\pm$ .22 \\
 & EEGPT & .793 $\pm$ .03 & .470 $\pm$ .04 & .735 $\pm$ .03 & .702 $\pm$ .10 & .658 $\pm$ .04 & .746 $\pm$ .05 \\
 & \textbf{Ours} & \textbf{.848 $\pm$ .06} & \textbf{.659 $\pm$ .02} & \textbf{.833 $\pm$ .06} & \textbf{.840 $\pm$ .03} & \textbf{.832 $\pm$ .05} & \textbf{.890 $\pm$ .04} \\
\midrule
\multirow{4}{*}{site5-EO} & DE+MLP & .738 $\pm$ .03 & .526 $\pm$ .01 & .759 $\pm$ .01 & .738 $\pm$ .03 & .731 $\pm$ .03 & .815 $\pm$ .04 \\
 & LaBraM & .779 $\pm$ .18 & .496 $\pm$ .04 & .736 $\pm$ .19 & .780 $\pm$ .23 & .754 $\pm$ .20 & .798 $\pm$ .20 \\
 & EEGPT & .724 $\pm$ .04 & .472 $\pm$ .04 & .704 $\pm$ .04 & .661 $\pm$ .05 & .664 $\pm$ .04 & .708 $\pm$ .05 \\
 & \textbf{Ours} & \textbf{.816 $\pm$ .05} & \textbf{.563 $\pm$ .08} & \textbf{.815 $\pm$ .03} & \textbf{.822 $\pm$ .04} & \textbf{.817 $\pm$ .05} & \textbf{.875 $\pm$ .03} \\
\midrule
\multirow{4}{*}{site5-EC} & DE+MLP & .729 $\pm$ .05 & .493 $\pm$ .03 & .748 $\pm$ .05 & .730 $\pm$ .04 & .725 $\pm$ .05 & .807 $\pm$ .04 \\
 & LaBraM & .774 $\pm$ .16 & .465 $\pm$ .06 & .761 $\pm$ .19 & .752 $\pm$ .16 & .755 $\pm$ .17 & .785 $\pm$ .18 \\
 & EEGPT & .726 $\pm$ .04 & .500 $\pm$ .05 & .693 $\pm$ .04 & .652 $\pm$ .03 & .641 $\pm$ .03 & .684 $\pm$ .04 \\
 & \textbf{Ours} & \textbf{.802 $\pm$ .06} & \textbf{.671 $\pm$ .08} & \textbf{.818 $\pm$ .05} & \textbf{.785 $\pm$ .09} & \textbf{.799 $\pm$ .07} & \textbf{.851 $\pm$ .05} \\
\midrule
\multirow{2}{*}{Average} & Best Baseline & .800 & .527 & .790 & .786 & .783 & .835 \\
 & \textbf{Ours} & \textbf{.824} & \textbf{.634} & \textbf{.816} & \textbf{.821} & \textbf{.815} & \textbf{.871} \\
\bottomrule
\end{tabular}
\end{sc}
\end{small}
\end{table*}

\begin{figure}[htbp]
\centering
\includegraphics[width=\columnwidth]{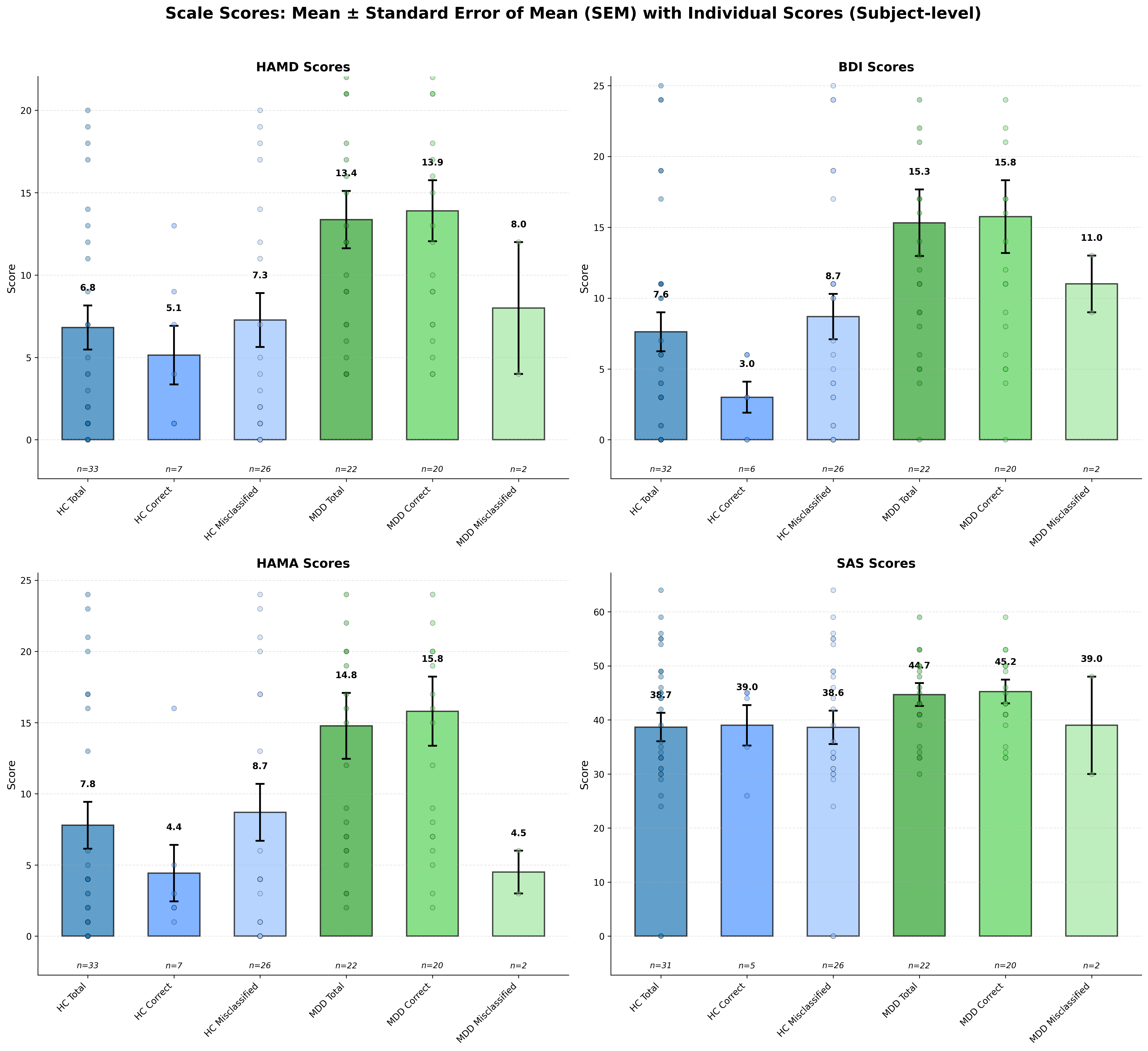} 
\caption{\textbf{Scales values of the subjects}The bars are color-coded, with blue representing HC and green representing MDD patients. For each group, the bars from left to right indicate: (i) the mean score for that category, (ii) the mean score of correctly classified subjects, and (iii) the mean score of misclassified subjects. The y-axis denotes the respective scale scores across four diagnostic dimensions: HAMD and BDI, and HAMA and SAS (Anxiety).}
\label{fig:error_subject}
\end{figure}

\textbf{Comparison with State-of-the-art.} As shown in Table~\ref{table:main_results}, our framework consistently outperforms existing methods across all experimental conditions. Compared to current state-of-the-art (SOTA) foundation models and the DE baseline(DE+MLP), our approach achieves an average improvement of \textbf{2.38\%--3.58\%} across standard metrics (B-Acc, Precision, Recall, F1, and AUROC). Most notably, in the challenging cross-site zero-shot scenarios, our model surpasses SOTA by a substantial margin of \textbf{10.7 percentage points}. These results underscore the effectiveness of our site-agnostic architecture for robust clinical screening.\\To further investigate the model's decision boundaries, we conducted a detailed analysis of misclassified subjects in relation to their clinical symptom severity (e.g., HAMD scores). As illustrated in Figure~\ref{fig:error_subject}, a clear \textbf{classification label limitation problem} is observed for misclassified individuals: healthy controls (HC) who were incorrectly predicted as MDD generally exhibited higher clinical scale scores compared to the correctly classified HC group. Conversely, MDD patients misclassified as HC tended to have significantly lower symptom severity scores. We do a further interpretability analysis to see what the models paying attention in Fig.\ref{fig:total_interpretability} and Appendix \ref{appendix:Interpretability}.\\

\textbf{The surprisingly high performance of the baseline mirrors classic pitfalls in data analysis.}The DE+MLP combination exhibits exceptionally high performance on Site 1; however, our analysis in Fig\ref{fig:de_mlp} and Appendix \ref{appendix:shortcut} demonstrates that this is a case of classic shortcut learning. Overall, \textbf{the model extensively exploits inherent differences in data distribution. Guided by the cross-entropy loss, it amplifies shortcuts that leverage site-specific characteristics to inflate accuracy—a pitfall that our research explicitly aims to figure out.}

\textbf{Analysis of Site-Invariance and Generalizability.} To further investigate the correlation between site-adversarial training and model generalization, we analyzed the performance fluctuations on Site 5 throughout the 10-fold cross-validation process. We monitored the HC/MDD site-discriminator accuracies alongside zero-shot performance. Results in table \ref{tab:discriminator_results} align with our hypothesis: strengthening the model's ability to suppress site-specific artifacts (i.e., decreasing site-discriminator accuracy) directly enhances its cross-site generalization. However, the site-discriminators maintained a relatively stable state in the late training stages, suggesting that our GRL-based approach effectively stabilizes the learning of site-invariant representations. Sensitivity analysis regarding domain adversarial intensity is detailed in Appendix \ref{appendix:domain_adv}.

\subsection{Impact of Site Diversity on Generalization}
\begin{figure}[ht]
\centering
\includegraphics[width=\columnwidth]{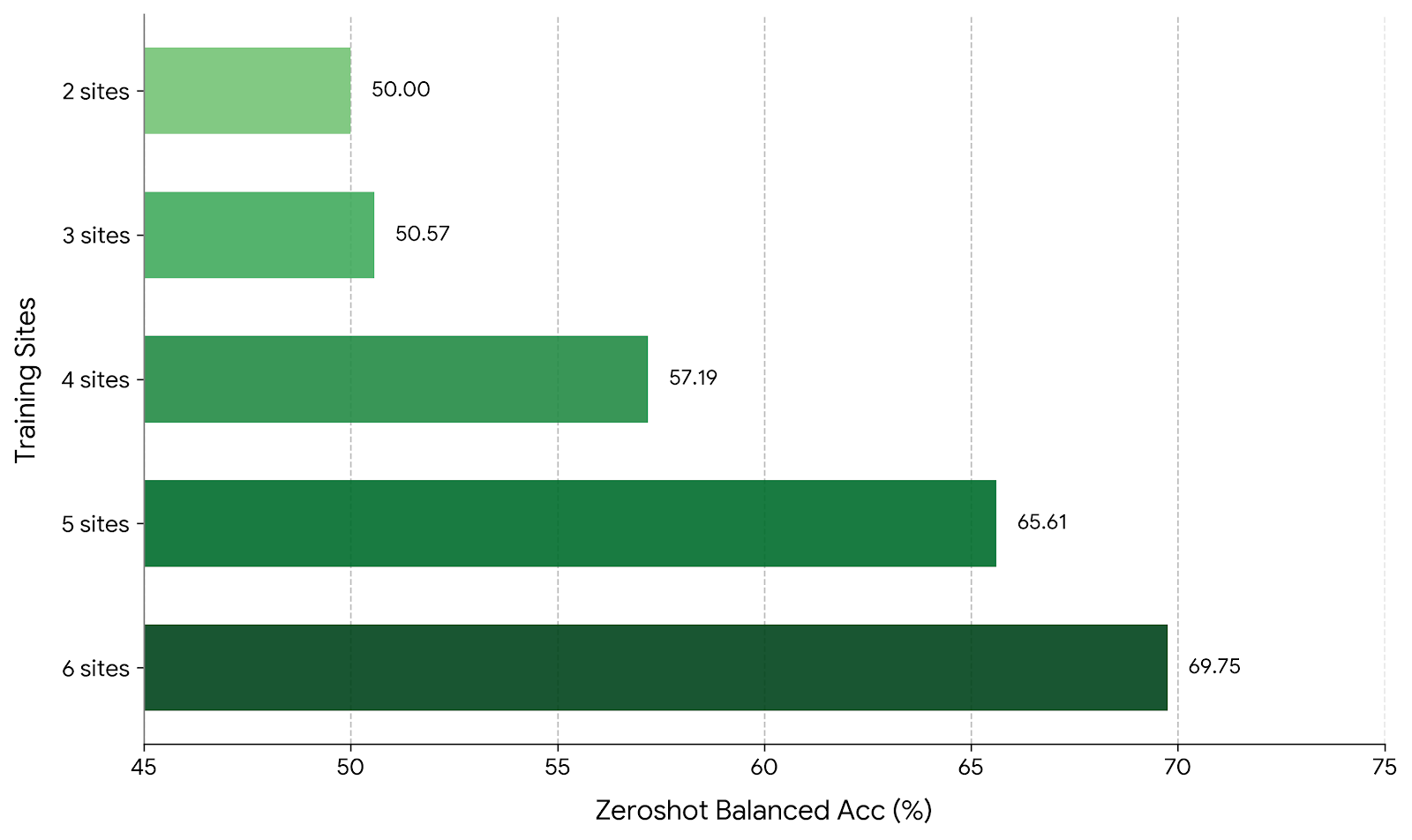} 
\caption{\textbf{Zero-shot generalization performance across varying training site scales.} The trend demonstrates a clear \textit{scaling law}: the balanced accuracy on the unseen test set increases monotonically as the number of training sites grows from 2 to 6. The model achieves peak performance (69.75\%) with 6 sites, underscoring the benefits of data diversity in enhancing cross-site robustness.}
\label{fig:scaling_law}
\end{figure}

To investigate how the breadth of training distributions influences model robustness, we conducted a scaling analysis by fine-tuning the model on an increasing number of sites, using Site 1 as the held-out target dataset. As illustrated in Figure~\ref{fig:scaling_law}, the model's performance exhibits a consistent upward trajectory as additional sites are integrated into the training phase. 
This observed scaling effect suggests that our architecture effectively leverages the biological variance present across different recording environments. Rather than over-parameterizing the model, the increase in site diversity acts as a powerful regularizer, enabling the framework to distill more universal and domain-invariant neural signatures.

\subsection{Ablation Studies}

To evaluate the individual contributions of our proposed components, we conducted systematic ablation experiments on Site 1 (Eyes-Open condition). We incrementally integrated: (i) multi-dataset generative pre-training, (ii) the Cross-Subject and Cross-Site (CSCS) contrastive loss, and (iii) the group-specific site-adversarial mechanism. 

\begin{table*}[t]
\centering
\caption{Ablation Studies on Site 1 (EO). We evaluate: (1) Generative Pre-training, (2) Cross subject and cross site Loss Design, and (3) site adv $L$.  indicates the best performance in each category.}
\label{table:ablation_all}
\vskip 0.15in
\begin{small}
\begin{sc}
\begin{tabular}{lcccccc}
\toprule
Configuration & B-Acc & ZS-BAcc & Precision & Recall & F1 & AUROC \\
\midrule
DE+MLP & .819 $\pm$ .06 & .570 $\pm$ .02 & \textbf{.815 $\pm$ .06} & .819 $\pm$ .06 & .810 $\pm$ .06 & \textbf{.874 $\pm$ .07} \\
\midrule
\textit{I. Pre-training} \\
w/o Pre-training & .545 $\pm$ .07 & .471 $\pm$ .09 & .500 $\pm$ .09 & .597 $\pm$ .08 & .540 $\pm$ .07 & .524 $\pm$ .10 \\
\textbf{With Pre-training} & \textbf{.832 $\pm$ .03} & \textbf{.645 $\pm$ .04} & .799 $\pm$ .07 & \textbf{.838 $\pm$ .07} & \textbf{.815 $\pm$ .04} & .870 $\pm$ .04 \\
\midrule
\textit{II. Loss \& CSCS Design} \\
w/o $\mathcal{L}$ (Baseline) & .646 $\pm$ .04 & .530 $\pm$ .06 & .584 $\pm$ .10 & .764 $\pm$ .10 & .650 $\pm$ .04 & .674 $\pm$ .05 \\
w/ $\mathcal{L}$ (No Cross-site) & \textbf{.847 $\pm$ .04} & .604 $\pm$ .02 & \textbf{.820 $\pm$ .06} & \textbf{.852 $\pm$ .05} & \textbf{.834 $\pm$ .04} & \textbf{.881 $\pm$ .05} \\
\textbf{Ours (Full)} & .832 $\pm$ .03 & \textbf{.645 $\pm$ .04} & .799 $\pm$ .07 & .838 $\pm$ .07 & .815 $\pm$ .04 & .870 $\pm$ .04 \\
\midrule
\textit{III. site adv $L$} \\
w/o $L$ & .728 $\pm$ .12 & .529 $\pm$ .08 & .663 $\pm$ .12 & \textbf{.859 $\pm$ .10} & .742 $\pm$ .09 & .763 $\pm$ .14 \\
\textbf{Ours (Full)} & \textbf{.832 $\pm$ .03} & \textbf{.645 $\pm$ .04} & \textbf{.799 $\pm$ .07} & .838 $\pm$ .07 & \textbf{.815 $\pm$ .04} & \textbf{.870 $\pm$ .04} \\
\bottomrule
\end{tabular}
\end{sc}
\end{small}
\end{table*}

To systematically evaluate the contribution of each proposed component, we conducted three sets of ablation experiments on Site 1 (EO), as summarized in Table~\ref{table:ablation_all}. 

\textbf{Effect of Generative Pre-training.} The first set of experiments underscores the critical role of our multi-dataset pre-training stage. Without pre-training (\textit{w/o Pre-training}), the model struggles to achieve competitive results, with the balanced accuracy dropping significantly to 0.545. This suggests that the high-dimensional noise inherent in raw EEG signals poses a substantial challenge for supervised fine-tuning alone.

\textbf{Impact of CSCS Loss Design.} We further examined the necessity of our Cross-Subject and Cross-Site contrastive loss and the site-adversarial mechanism ($\mathcal{L}$). A compelling observation arises: while the model \textit{without} the cross-site sampling design achieves a slightly higher in-site balanced accuracy compared to our full model (0.847 vs. 0.832), its zero-shot performance (\textit{ZS-BAcc}) suffers a marked decline to 0.604. This phenomenon confirms a pivotal trade-off in clinical EEG modeling. By intentionally incorporating cross-site variance, our full framework prioritizes the distillation of \textbf{domain-invariant neural signatures}, sacrificing a marginal amount of local accuracy to achieve superior robustness in unseen environments.

\textbf{Contribution of Site-adversarial Loss} Finally, we ablated the site-specific module. The results demonstrate that the inclusion of the discriminators is instrumental for stabilizing the latent space. Removing the adversarial loss leads to a performance drop across nearly all metrics, particularly reducing the B-Acc from 0.832 to 0.728. This highlights that the  discriminators effectively functions as a robust feature filter, bridging the gap between real neural representation and clinical category knowledge.

In summary, the ablation studies validate that the synergy of generative priors, site-invariant constraints, and the specialized architecture of site-adversarial module is essential for overcoming the distribution shifts in multi-center clinical diagnostics.

\section{Discussion and conclusion}
In this study, we proposed a robust cross-subject and cross-site EEG classification framework for MDD and HC based on contrastive concept. Utilizing our self-constructed multi-site diagnostic dataset, we demonstrated the errors which limits the cross-site generalization capability. Our findings elucidate this formulation-driven approach for enhancing generalization performance represents a generalizable paradigm that could serve as a broad reference for future research in this field, offering a highly promising direction for future real-world clinical applications.

\textbf{Limitations and future directions}

 \textbf{We recognize that a self-collected dataset provides only a partial solution to the confounding factors of inter-device variability and observer bias,} while some studies indicate a Cohen’s kappa of only 0.28 for MDD. Consequently, the labels within any psychiatric dataset carry an inherent risk of being misleading. To some extent, this also suggests that pushing intra-dataset performance to a level of 0.9 or higher may be fundamentally unrealistic. In the current study, we remain unable to resolve the issues stemming from such inherent labeling errors.
 
\textbf{Furthermore, this study does not differentiate between the various stages of disease progression, such as the acute, maintenance, and consolidation phases.} The inclusion of patients in the consolidation phase—those who have fully resumed normal daily activities and whose treatment focus has shifted to preventing relapse—may introduce variability into the model’s evaluation. Although existing research has suggested that patients in the consolidation phase still exhibit "incomplete recovery" at the neural and vascular levels, future work should involve more granular staging of patients. By conducting stage-specific analysis,we aim to extract potential neurobiomarkers associated with clinical recovery.

\section*{Software and Data}
Consistent with the EMBARC protocol, our data can be made available to researchers upon the execution of a formal Data Use Agreement (DUA).
The code is available at:

\section*{Impact Statement}

This paper presents a rigorous framework for robust EEG representation learning and can be transfered to other neuroscience field with strong domain shifts —a formidable barrier that currently prevents most EEG-based deep learning models from advancing into real-world clinical deployment. We recognize that the catastrophic performance drop of neural decoding models in unseen environments often stems from a reliance on site-specific artifacts rather than authentic neural signatures.

By formulating domain bias through 3 factors, our work systematically addresses these confounding factors through integrated strategies: from meticulous data cohort design to noise-resilient algorithmic architectures. To some extent, we do not propose a new loss, but a principled decomposition of domain bias that constrains how losses should be designed and interpreted. Through our experiments, we have replicated some of these errors and the resulting false positives, which allows us to shift our focus more effectively toward the interpretation of the performance.

Our approach moves beyond pursuing marginal gains in accuracy on closed datasets, instead prioritizing the cross-site generalization and robustness essential for clinical reliability.

We envision that our methodology, the proposed CRCC framework, and the high-standard multi-site database will provide a benchmark and a generalizable philosophy for the community. This work aims to empower future researchers to develop AI-driven biomarkers that are truly clinically actionable, ultimately facilitating the large-scale screening and objective diagnosis of major depressive disorder (MDD) in diverse, real-world populations.

\nocite{langley00}

\bibliography{example_paper}
\bibliographystyle{icml2026}

\newpage
\appendix
\label{appendix:paradigm}
\onecolumn
\section{EEG paradigm and pre-processing pipeline}
Prior to the experiment, each participant completed the Beck Depression Inventory (BDI) and the Self-rating Anxiety Scale (SAS). This was followed by semi-structured interviews conducted by certified evaluators, who had passed inter-rater reliability assessments, to complete the Hamilton Depression Rating Scale (HAMD) and the Hamilton Anxiety Rating Scale (HAMA). The entire clinical assessment process was estimated to last 15--20 minutes.\cite{yan2025differences}

Each participant underwent a sequence of resting-state recordings: 2 minutes with eyes open, followed by 2 minutes with eyes closed. This sequence was then repeated to obtain two full cycles (open–closed–open–closed), ensuring sufficient relaxation and physiological stabilization. Participants were instructed to press a button to advance to each subsequent stage, and an auditory cue was presented to signal the end of each eyes-closed interval.

EEG signals were recorded using a wireless EEG acquisition system (NeuSen.W32) at a sampling rate of 1000 Hz. Thirty-two saline-based sponge electrodes were positioned according to the international 10--20 system. Electrode impedance was maintained below 30~k\(\Omega\) throughout the experiment to ensure high-quality signal acquisition.

The EEG preprocessing pipeline was implemented using the MNE-python library. First, A1/A2 electrodes were excluded prior to executing standard pre-processing steps. Subsequently, signal clipping, band-pass filtering (0.05--75~Hz), notch filtering, and downsampling (to 250~Hz) were performed, followed by the application of a common average reference (CAR).

\section{Hyperparameter of the model}
\label{appendix:hyperparams}
The hyperparameters in model pre-training and fine-tuning are shown in Table\ref{table:appendix_hyperparams} including parameters of the encoder and the fine-tuning layer, also its optimizer design.
As detailed in Table \ref{table:appendix_hyperparams}, we provide a comprehensive list of the hyperparameters used during the multi-dataset pre-training and the subsequent cross-site fine-tuning (SFT) stages. 

\begin{table}[ht]
\centering
\large 
\caption{Hyperparameter configurations for both Pre-training and Fine-tuning stages.}
\label{table:appendix_hyperparams}
\vskip 0.15in
\renewcommand{\arraystretch}{1.2} 
\begin{tabular}{l c c}
\toprule
\textbf{Category \& Parameter} & \textbf{Pre-training} & \textbf{Fine-tuning (SFT)} \\
\midrule
\rowcolor[gray]{0.95} \textit{Model Architecture} & & \\
Projection Dimension & --- & 8 \\
Drop Path Rate & --- & 0.1 \\
Mask Ratio & 0.5 & --- \\
Number of Classes & --- & 2 \\

\midrule
\rowcolor[gray]{0.95} \textit{Optimization} & & \\
Optimizer & AdamW & AdamW \\
Base Learning Rate & $5 \times 10^{-4}$ & $5 \times 10^{-4}$ \\
Adam $\beta$ Coefficients & (0.9, 0.98) & (0.9, 0.999) \\
Weight Decay & 0.05 & 0.05 \\
LLRD\textsuperscript{*} & --- & 0.65 \\

\midrule
\rowcolor[gray]{0.95} \textit{Training Schedule} & & \\
Batch Size & 512 & 512 \\
Total Epochs & 30 & 50 \\
Warmup Epochs & 3 & 3 \\
Loss Coefficient ($\alpha$ / $\beta$) & $\beta = 1.0$ & $\alpha = 0.6$ \\

\midrule
\rowcolor[gray]{0.95} \textit{System \& Hardware} & & \\
Random Seed & 12345 & 12345 \\
Accumulate Grad Batches & 1 & 1 \\
Precision & FP32 & FP32 \\
\bottomrule
\end{tabular}
\begin{flushleft}
\small \textsuperscript{*}LLRD: Layer-wise Learning Rate Decay.
\end{flushleft}
\end{table}

\section{Datasets used in the pretraining}
\label{appendix:pretrain}
Both our proposed model and the baseline foundation models were pretrained on a consistent set of EEG datasets. The specific open-source datasets used are as follows:

\textbf{SEED DATASET: }\url{https://bcmi.sjtu.edu.cn/home/seed/#}\\
\textbf{LEMON: } \url{https://fcon_1000.projects.nitrc.org/indi/retro/MPI_LEMON.html} \\
\textbf{TUHEEG: }\url{https://isip.piconepress.com/projects/nedc/html/tuh_eeg/index.shtml}\\
\textbf{Childhood Sexual Abuse and problem drinking in women: Neurobehavioral mechanisms: }\url{https://openneuro.org/datasets/ds003602/versions/1.0.1}\\
\textbf{E-CAM-S: }\url{https://bdsp.io/content/e-cam-s/1.0/}\\
\textbf{Predict: }\url{https://openneuro.org/datasets/ds005114/versions/1.0.0}, \url{https://predict.cs.unm.edu/downloads.php}\\
\textbf{Resting-state EEG data before and after cognitive activity across the adult lifespan and a 5-year follow-up: }\url{https://openneuro.org/datasets/ds005385/versions/1.0.0}\\
\textbf{FACED: }\url{https://www.synapse.org/Synapse:syn50614194/discussion/default}\\
\textbf{One-person-microstate: }\url{https://www.nature.com/articles/s41597-024-03241-z}

\section{Interpretability of the Model}
\label{appendix:Interpretability}
In this section, we delve into the internal representations of the SFT model to demystify what knowledge it has acquired during training. The following figure \ref{fig:eegpt_four_plots} illustrates the model's attention weights (or input preferences) for EEGPT and our model after excluding Site 1 or Site 5 (eyes close condition).

\begin{figure}[t]
     \centering
     \begin{subfigure}[b]{0.4\textwidth}
         \centering
         \includegraphics[width=\textwidth]{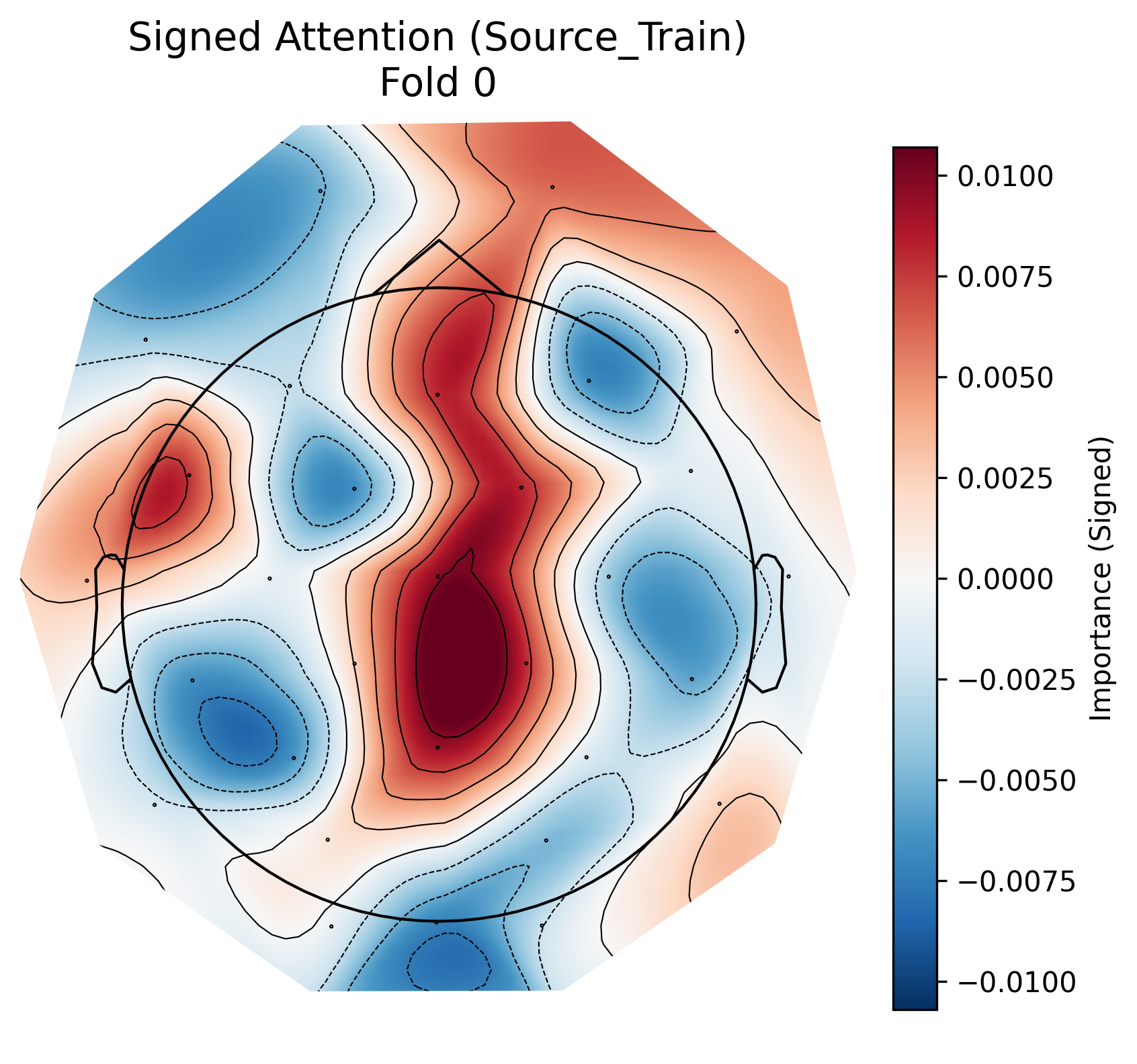}
         \caption{The localization and weight distribution of key discriminative regions identified by EEGPT, with the exception of Site 1 (EC)}
         \label{fig:image1}
     \end{subfigure}
     \hfill
     \begin{subfigure}[b]{0.4\textwidth}
         \centering
         \includegraphics[width=\textwidth]{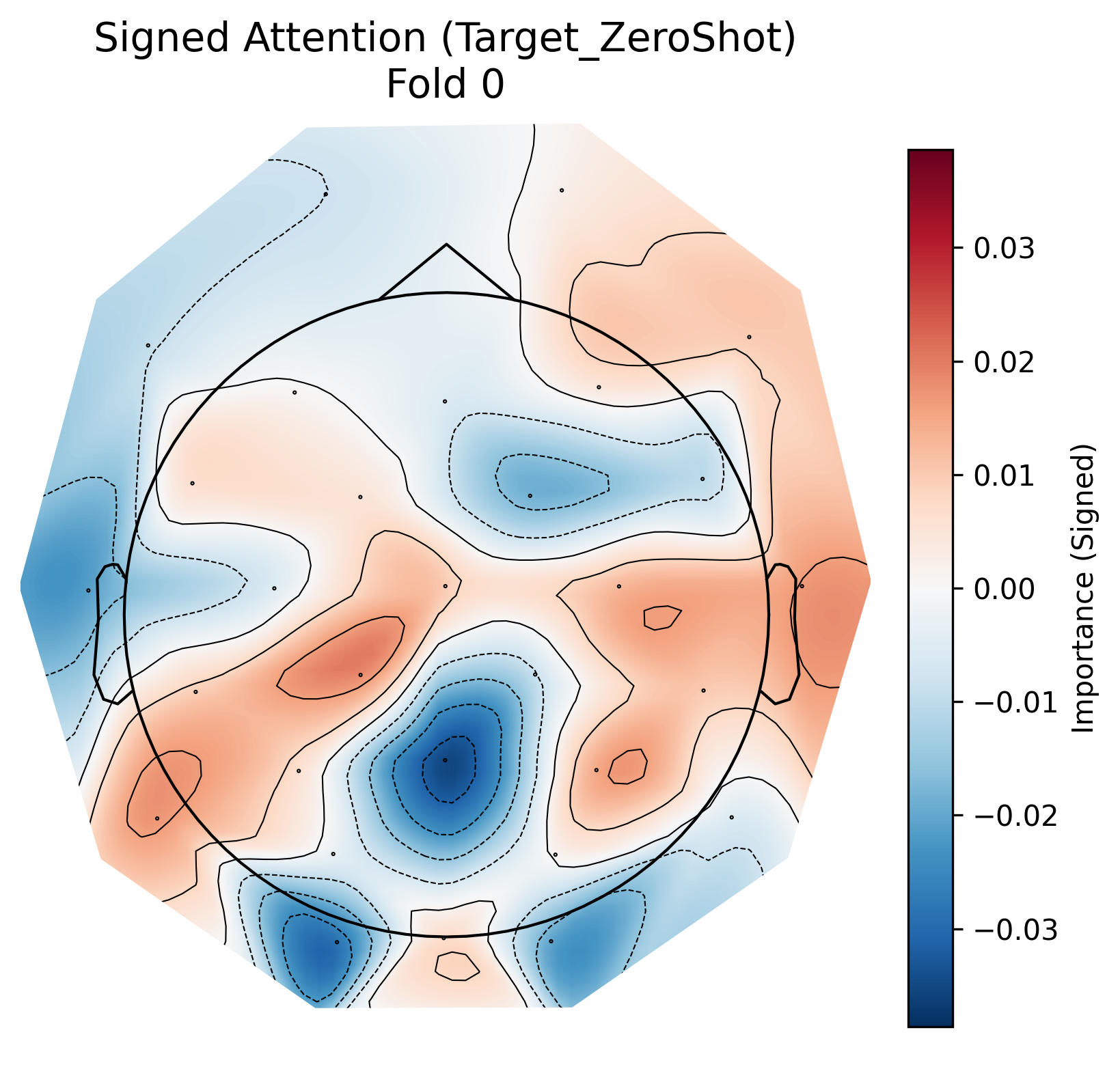}
         \caption{Weight analysis of EEGPT's decision-making process under Site 1 (EC), illustrating the specific data features upon which the model's final judgment relies.}
         \label{fig:image2}
     \end{subfigure}

     \vspace{0.8em} 

     \begin{subfigure}[b]{0.4\textwidth}
         \centering
         \includegraphics[width=\textwidth]{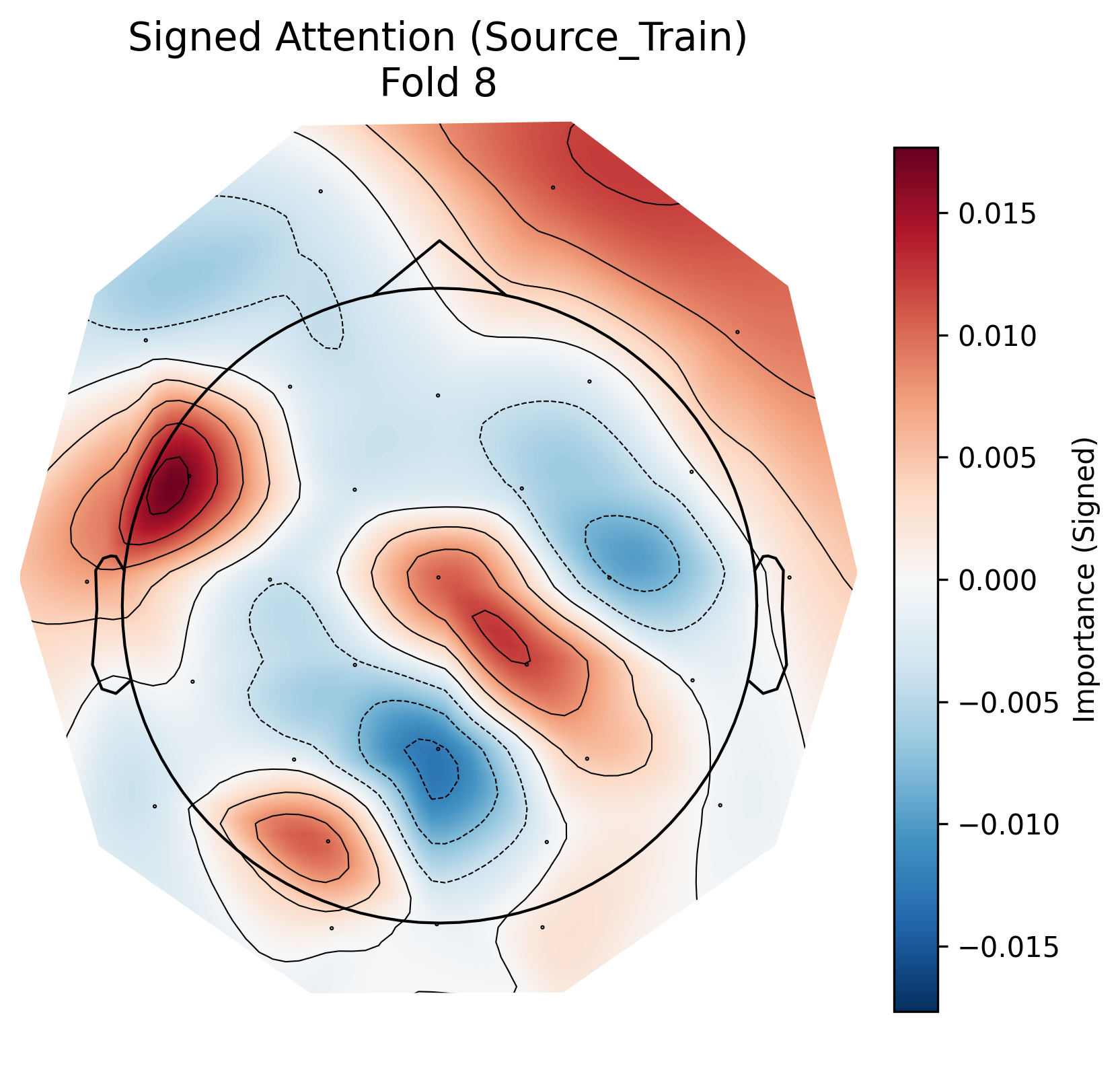}
         \caption{The localization and weight distribution of key discriminative regions identified by EEGPT, with the exception of Site 5 (EC)}
         \label{fig:image3}
     \end{subfigure}
     \hfill
     \begin{subfigure}[b]{0.4\textwidth}
         \centering
         \includegraphics[width=\textwidth]{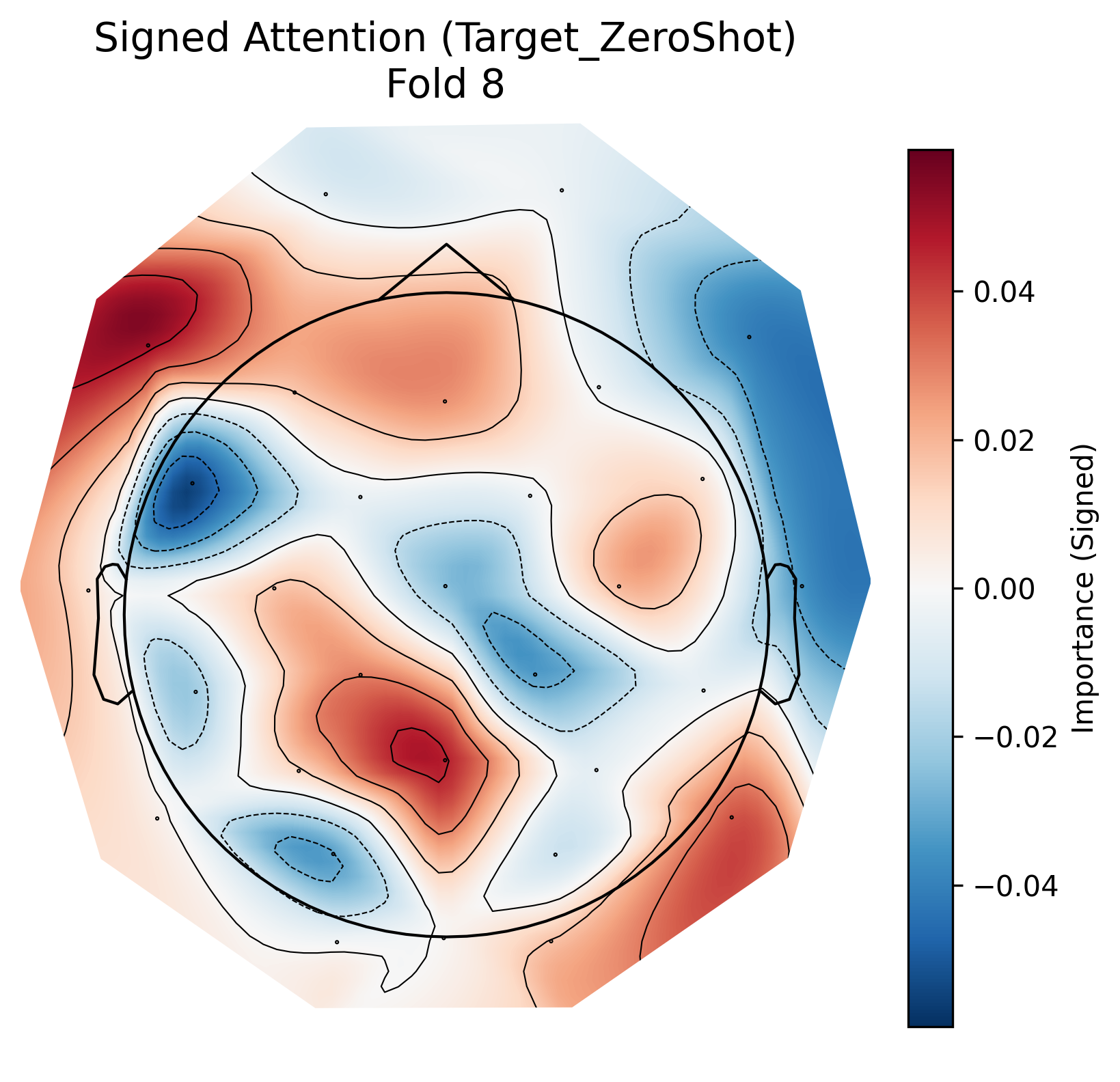}
         \caption{Weight analysis of EEGPT's decision-making process under Site 5 (EC), illustrating the specific data features upon which the model's final judgment relies.}
         \label{fig:image4}
     \end{subfigure}

     \caption{Comparative Interpretability Analysis of EEGPT: Intuitively, this explores the basis of the model's classification decisions. Greater alignment between the focus of training data and generalization data indicates that their distributions are more similar from the model's perspective, thereby leading to superior generalization performance.}
     \label{fig:eegpt_four_plots}
\end{figure}

\begin{figure}[htbp]
    \centering
    
    \begin{subfigure}{\textwidth}
        \centering
        \includegraphics[width=0.95\textwidth]{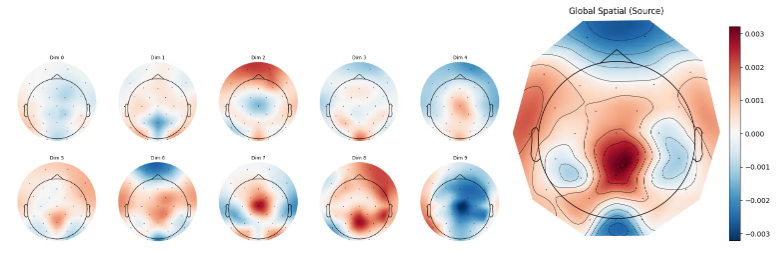} 
        \caption{Interpretability analysis of CRCC for Trained data except Site 1 (EC): Localization of key discriminative regions.}
        \label{fig:site2_analysis}
    \end{subfigure}
    
    \vspace{1em} 
    
    \begin{subfigure}{\textwidth}
        \centering
        \includegraphics[width=0.95\textwidth]{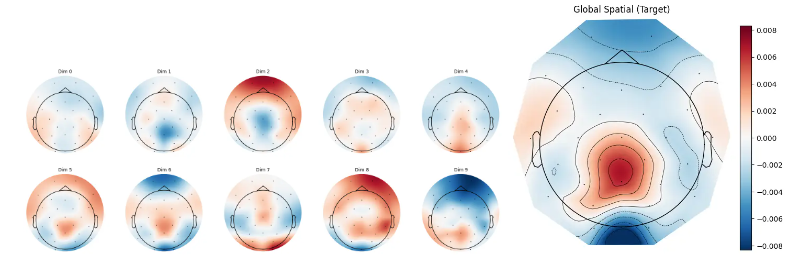}
        \caption{Zero-shot generalization of CRCC on site 1 (EC): Attention alignment on unseen domains.}
        \label{fig:site6_analysis}
    \end{subfigure}
    
    \vspace{1.em}
    
    \begin{subfigure}{\textwidth}
        \centering
        \includegraphics[width=0.95\textwidth]{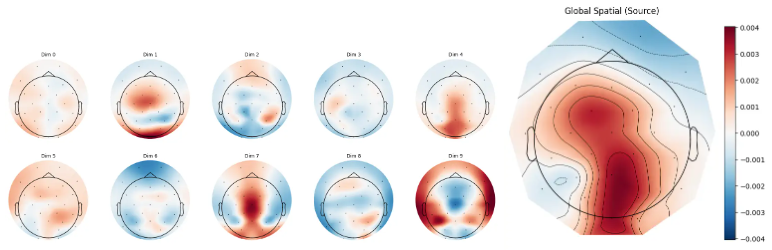}
        \caption{Interpretability analysis of CRCC for Trained data except Site 5 (EC): Localization of key discriminative regions.}
        \label{fig:zeroshot_alignment}
    \end{subfigure}
    
    \vspace{1.em}
    
    \begin{subfigure}{\textwidth}
        \centering
        \includegraphics[width=0.95\textwidth]{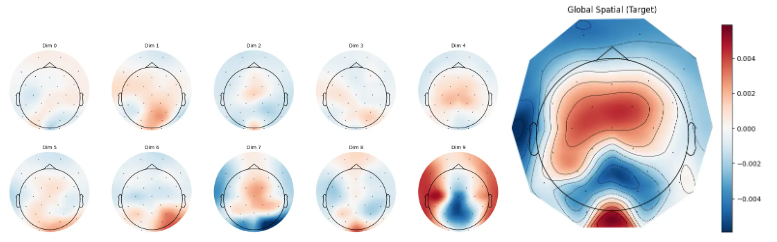}
        \caption{Zero-shot generalization of CRCC on site 5 (EC): Attention alignment on unseen domains..}
        \label{fig:feature_comparison}
    \end{subfigure}

    \caption{Multi-level interpretability and consistency analysis of CRCC(ours) across various sites and generalization scenarios.}
    \label{fig:total_interpretability}
\end{figure}

To transparently uncover the neural mechanisms learned by our model and rigorously validate its generalization capability across unseen domains (e.g., zero-shot subjects), we propose a fine-grained interpretability framework. Unlike traditional methods that focus solely on spatial topology, our approach employs a \textit{dual-view attribution analysis} to disentangle the contributions of spatial brain regions and latent feature dimensions.

\subsection{Channel-Feature Dual Attribution Extraction}
We utilize Integrated Gradients (IG), a strictly axiomatic attribution method, to quantify the contribution of input features to the model's decision-making. Let $F: \mathbb{R}^{C \times D} \rightarrow [0, 1]$ denote the mapping function of our pre-trained model, where $C$ is the number of EEG channels and $D$ represents the dimensionality of the latent features (or temporal segments).
For a given input sample $\mathbf{X}$ and a zero baseline $\mathbf{X}'$, we first compute the comprehensive Attribution Matrix $\mathbf{M} \in \mathbb{R}^{C \times D}$. The element $M_{c,d}$, representing the importance of the $d$-th feature in the $c$-th channel, is calculated as:

\begin{equation}
M_{c,d}(\mathbf{X}) = (\mathbf{X}_{c,d} - \mathbf{X}'_{c,d}) \times \int_{\alpha=0}^{1} \frac{\partial F(\mathbf{X}' + \alpha \times (\mathbf{X} - \mathbf{X}'))}{\partial \mathbf{X}_{c,d}} d\alpha
\end{equation}

In practice, this integral is approximated using a Riemann sum with $m=20$ steps to ensure convergence. 
To interpret this high-dimensional matrix, we perform marginalization along the feature and channel axes, yielding two distinct importance vectors:

\begin{itemize}
    \item \textbf{Global Spatial Importance ($\mathbf{v}_{spa} \in \mathbb{R}^C$):} By averaging $\mathbf{M}$ over the feature dimension, we identify the most salient brain regions (e.g., frontal or temporal lobes) utilized by the model:
    \begin{equation}
    \mathbf{v}_{spa}[c] = \frac{1}{D} \sum_{d=1}^{D} M_{c,d}
    \end{equation}
    
    \item \textbf{Global Feature Importance ($\mathbf{v}_{feat} \in \mathbb{R}^D$):} By averaging $\mathbf{M}$ over the channel dimension, we isolate the specific latent features that drive the prediction, independent of spatial location:
    \begin{equation}
    \mathbf{v}_{feat}[d] = \frac{1}{C} \sum_{c=1}^{C} M_{c,d}
    \end{equation}
\end{itemize}

This decomposition allows us to verify whether the model learns specific, biologically plausible feature patterns rather than relying on channel-specific artifacts.

\subsection{Canonical Alignment for Cross-Subject Aggregation}
A significant challenge in multi-dataset EEG analysis is the heterogeneity of electrode montages (i.e., varying channel numbers and coordinates across subjects). To enable cross-subject comparison, we map subject-specific attributions to a standardized canonical space (e.g., the standard 10-20 system).

Let $\mathcal{P}_{obs} = \{\mathbf{p}_1, \dots, \mathbf{p}_C\}$ be the 3D coordinates of the observed channels for a specific subject, and $\mathcal{P}_{can} = \{\hat{\mathbf{p}}_1, \dots, \hat{\mathbf{p}}_K\}$ be the coordinates of target canonical channels. We employ a nearest-neighbor mapping strategy where each canonical channel $\hat{k}$ inherits the attribution values from the spatially closest observed channel $c^*$:

\begin{equation}
c^* = \operatorname*{arg\,min}_{c \in \{1, \dots, C\}} || \hat{\mathbf{p}}_{\hat{k}} - \mathbf{p}_c ||_2
\end{equation}

The aligned importance matrix $\hat{\mathbf{M}}$ is constructed using these mapped channels, ensuring that the spatial topology is preserved and comparable across different datasets.

\subsection{Multi-Level Cross-Domain Consistency Quantification}
To quantitatively demonstrate that our model learns domain-invariant biomarkers rather than overfitting to domain-specific noise, we compare the attribution patterns between the source domain (Training Set) and the target domain (Zero-Shot Test Set). 

Let $\hat{\mathbf{M}}^{(S)}$ and $\hat{\mathbf{M}}^{(T)}$ denote the average aligned attribution matrices for the source and target domains, respectively. We propose a multi-level consistency evaluation using the Pearson Correlation Coefficient (PCC), denoted as $r$, at three levels of granularity:

\begin{enumerate}
    \item \textbf{Matrix-Level Consistency ($r_{mat}$):} This metric measures the overall structural similarity of the learned representations across the entire channel-feature space:
    \begin{equation}
    r_{mat} = \text{PCC}\left(\text{vec}(\hat{\mathbf{M}}^{(S)}), \text{vec}(\hat{\mathbf{M}}^{(T)})\right)
    \end{equation}
    where $\text{vec}(\cdot)$ denotes the vectorization operation.
    \item \textbf{Spatial Consistency ($r_{spa}$):} This metric evaluates whether the model consistently focuses on the same brain regions across domains, indicating robust spatial localization:
    \begin{equation}
    r_{spa} = \text{PCC}\left(\mathbf{v}_{spa}^{(S)}, \mathbf{v}_{spa}^{(T)}\right)
    \end{equation}  
    \item \textbf{Feature Consistency ($r_{feat}$):} This metric verifies if the critical latent features identified in the training set remain salient in the zero-shot test set:
    \begin{equation}
    r_{feat} = \text{PCC}\left(\mathbf{v}_{feat}^{(S)}, \mathbf{v}_{feat}^{(T)}\right)
    \end{equation}
\end{enumerate}

High correlation values ($r \approx 1$) across these metrics provide strong evidence that the model has successfully extracted generalized neural patterns that are robust to the domain shifts inherent in cross-subject EEG analysis.

In our model, the $r_{mat}$ (or $r$) for Site 1 (EC) reaches 0.8058 (while EEGPT is 0.6) and 0.27 (while EEGPT is -0.7) on site 5 (EC), indicating that the model effectively transfers knowledge from the training set to the zero-shot generalization test set and extracts robust domain-invariant features. Furthermore, the model assigns greater weight to prefrontal alpha signals (dim2) while placing less emphasis on occipital signals, a finding that aligns with the majority of clinical studies on MDD populations. This suggests that our model, to a certain extent, bypasses site-specific noise to extract authentic neural representations of the disorder.

\section{Contribution and Analysis of Site Adversarial Learning}
\label{appendix:domain_adv}
We evaluate the relationship between accuracy and generalization performance for the HC and MDD discriminators of the site adversarial functions.\\We selected both site 1 and 5 for further evaluation. We applied the weights from every epoch of the 10-fold cross-validation to this test set and recorded the accuracies of the HC and MDD discriminators on the validation set. The experimental results provide preliminary evidence for our hypothesis: enhancing the model’s site-invariant capability (the ability to remove site-specific information) improves its generalization. However, the fluctuations in this site-invariant capability were not pronounced; in fact, the site discriminators remained in a relatively stable state during the later stages of training.\\

\begin{table}[htbp]
\centering
\caption{discriminators acc and generalization ability}
\label{tab:discriminator_results}
\small 
\begin{tabular}{l|l|c|c|c|c}
\toprule
\textbf{Site} & \textbf{Epoch} & \textbf{Balanced Accuracy} & \textbf{Zeroshot BA} & \textbf{HC Discriminator Acc} & \textbf{MDD Discriminator Acc} \\ \midrule
\multirow{2}{*}{site1\_eo} & best\_in\_val  & \textbf{0.832 ± 0.034} & 0.645 ± 0.043 & \underline{0.640 ± 0.106} & \underline{0.669 ± 0.106} \\
                           & best\_in\_test & 0.683 ± 0.025          & \textbf{0.775 ± 0.060} & 0.641 ± 0.109          & 0.669 ± 0.107          \\ \hline
\multirow{2}{*}{site1\_ec} & best\_in\_val  & \textbf{0.848 ± 0.044} & 0.659 ± 0.018 & \underline{0.637 ± 0.101} & 0.660 ± 0.087          \\
                           & best\_in\_test & 0.680 ± 0.020          & \textbf{0.820 ± 0.040} & 0.648 ± 0.096          & \underline{0.658 ± 0.088} \\ \hline
\multirow{2}{*}{site5\_eo} & best\_in\_val  & \textbf{0.816 ± 0.040} & 0.563 ± 0.029 & 0.573 ± 0.063          & 0.493 ± 0.077          \\
                           & best\_in\_test & 0.609 ± 0.045          & \textbf{0.761 ± 0.063} & \underline{0.563 ± 0.089} & \underline{0.484 ± 0.073} \\ \hline
\multirow{2}{*}{site5\_ec} & best\_in\_val  & \textbf{0.802 ± 0.053} & 0.671 ± 0.029 & 0.587 ± 0.055          & 0.488 ± 0.150          \\
                           & best\_in\_test & 0.701 ± 0.025          & \textbf{0.770 ± 0.074} & \underline{0.583 ± 0.054} & \underline{0.478 ± 0.141} \\ \bottomrule
\end{tabular}
\end{table}

\begin{figure}[htbp]
    \centering
    \begin{subfigure}[b]{0.8\textwidth}
        \centering
        \includegraphics[width=\textwidth]{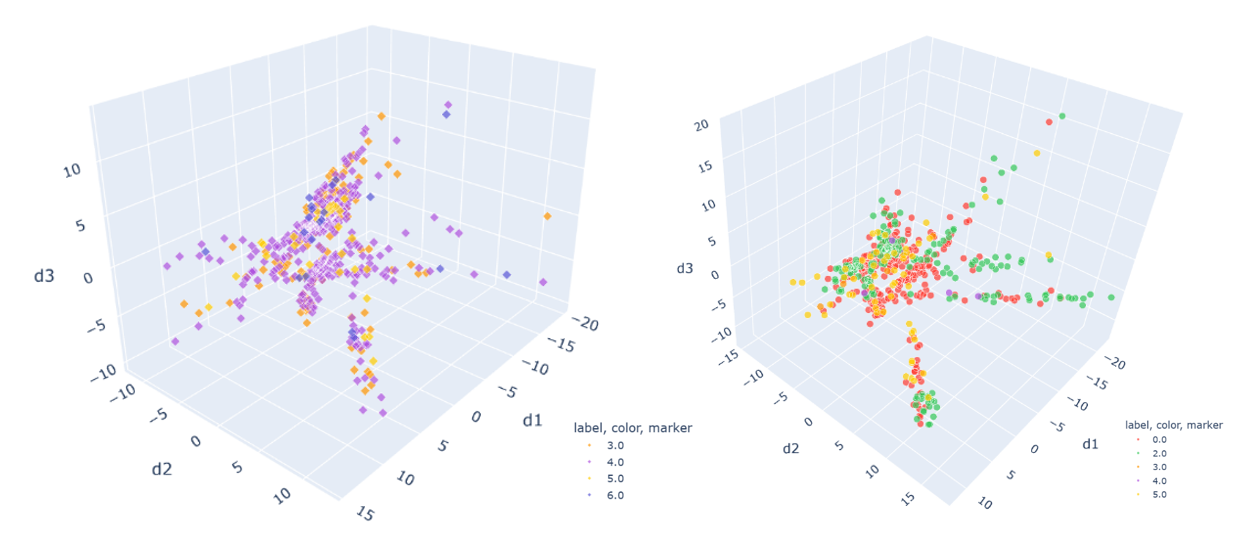} 
        \caption{EEGPT's Embedding}
        \label{fig:our_model}
    \end{subfigure}
    
    \vspace{1em} 

    \begin{subfigure}[b]{0.8\textwidth}
        \centering
        \includegraphics[width=\textwidth]{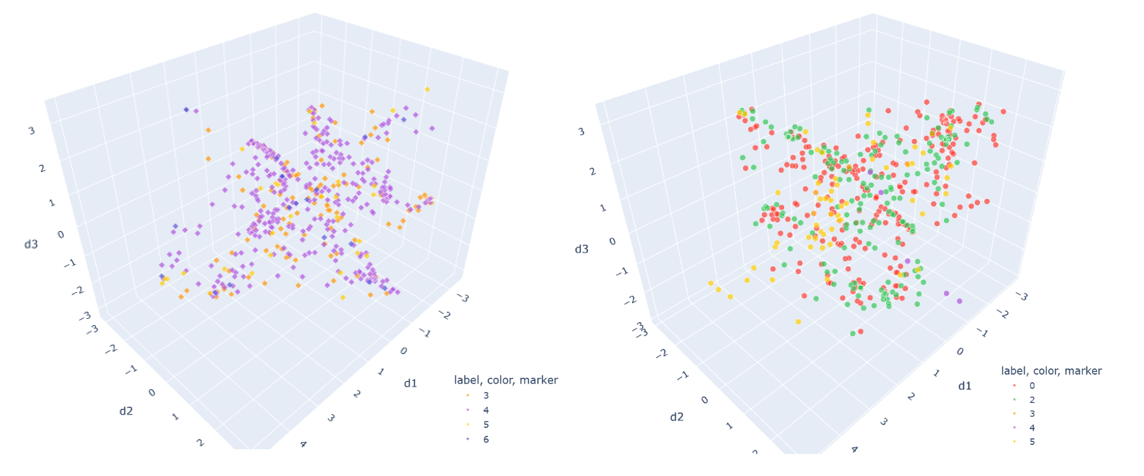} 
        \caption{Our Model's Embedding}
        \label{fig:eegpt}
    \end{subfigure}

    \caption{Comparison of latent space embeddings between EEGPT and our model. Our model exhibits lower clustering across different sites (represented by distinct colors), implying that data from a new site is more likely to follow a similarly broad distribution. This characteristic facilitates the reduction of site-specific noise, thereby enhancing the extraction of effective biomarkers. }
    \label{fig:combined_visualization}
    
\end{figure}

Furthermore, we visualize the embedding outputs from the layer preceding the output layer of our model alongside those from EEGPT to observe the mixing of data across different sites in Fig.\ref{fig:combined_visualization}. As expected, the data from various sites in our model exhibit a higher degree of overall mixing, which aligns more closely with the clinical reality of the disease—where subject features are broadly distributed throughout the latent space. In contrast, the embeddings from EEGPT show a degree of clustering, suggesting that the model relies to some extent on site-specific noise.

\section{Typical shortcut analysis of DE+MLP applied to Site 1}
\label{appendix:shortcut}
The surprisingly high performance of the DE+MLP baseline on Site 1 reflects a classic issue in dataset construction within this field: the separation of acquisition periods and scenarios between patients and healthy controls. As collectively illustrated in Figures 6, 7, and 8, different sites are distinctly separated in the visualization. This separation directly leads to inflated accuracy on the validation set when the dataset is composed of a single site. Furthermore, when the generalization dataset involves 'batch acquisition' or when healthy control recruitment fails to achieve demographic matching, the model learns substantial shortcuts. As shown in Figures 7 and 8, the MDD group from Site 1 significantly deviates from the data distribution of the validation set. Under the design of traditional cross-entropy loss functions, these shortcuts are further amplified. Consequently, the model fails to capture neuroactivities highly relevant to the disease, leading researchers to have overly optimistic expectations of the algorithm's performance. Therefore, the rigorous control design of Site 5, along with the model's zero-shot generalization capability on this site, serves as the most critical metric for distinguishing model performance.

\begin{figure}[htbp]
    \centering
    \begin{subfigure}[b]{0.8\textwidth}
        \centering
        \includegraphics[width=\textwidth]{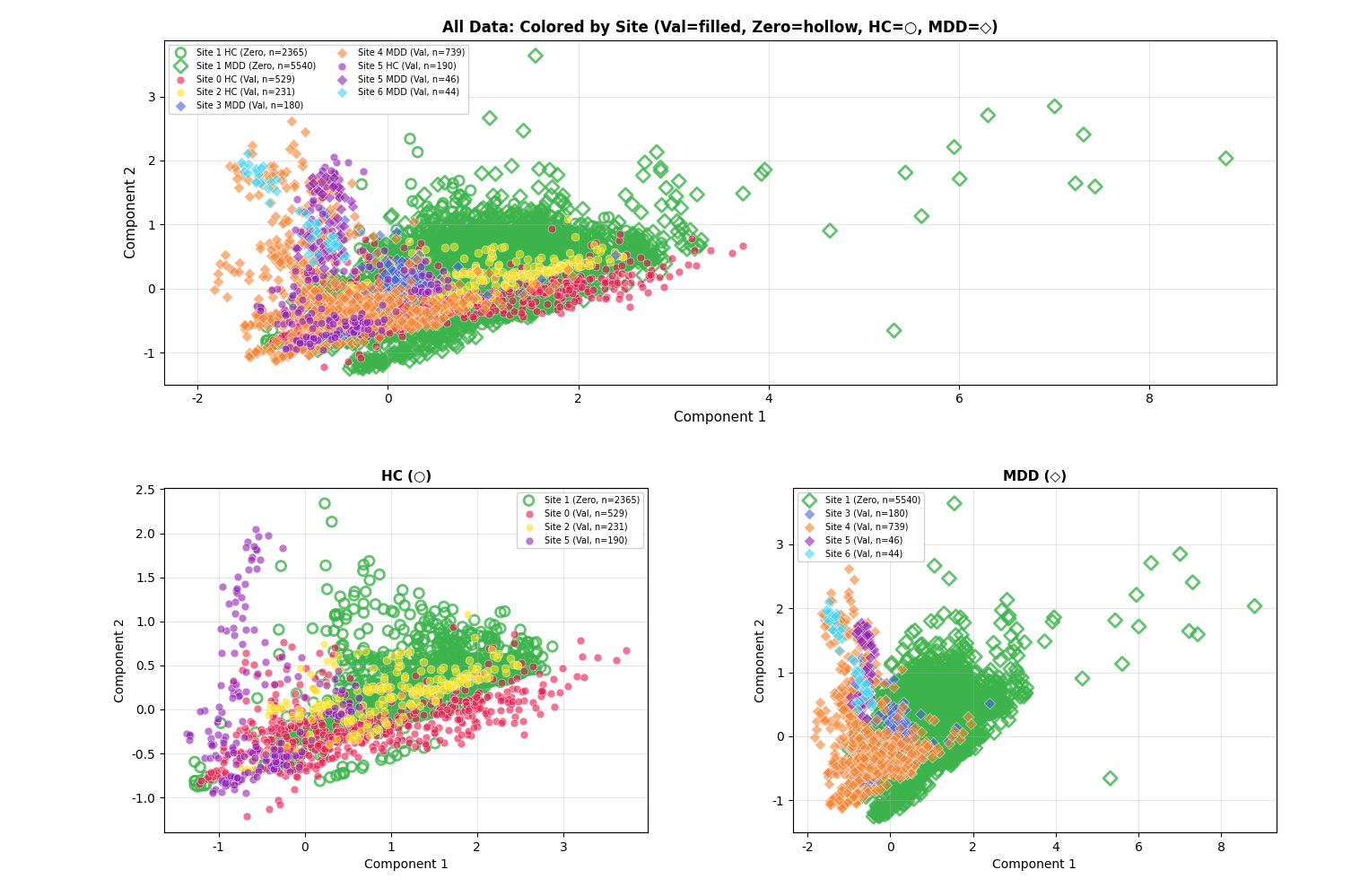} 
        \caption{one of the best validation performance fold of the DE+MLP model}
        \label{fig:de_mlp_1}
    \end{subfigure}
    
    \vspace{1em} 

    \begin{subfigure}[b]{0.8\textwidth}
        \centering
        \includegraphics[width=\textwidth]{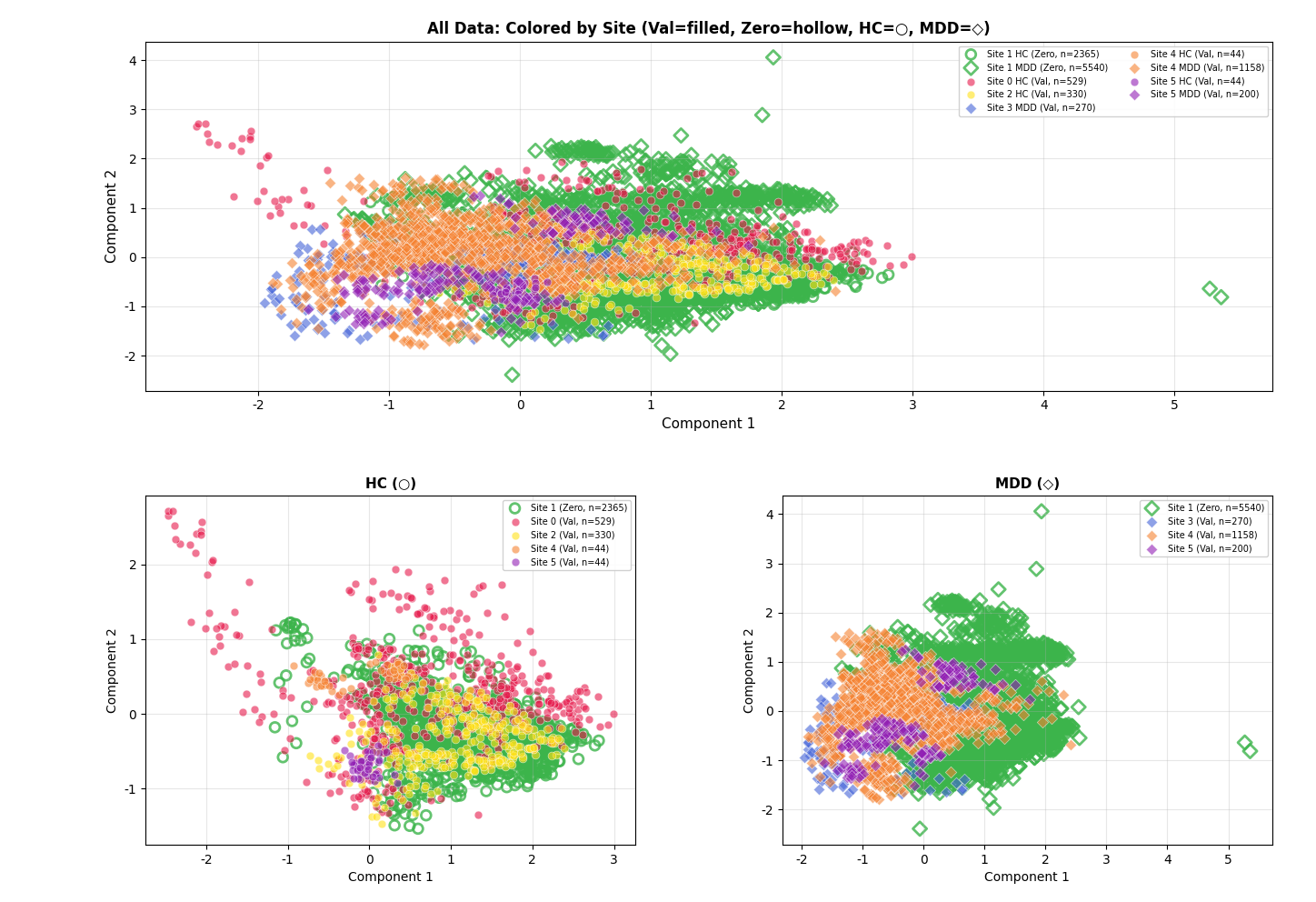} 
        \caption{one of the best validation performance fold of the DE+MLP model}
        \label{fig:de_mlp_2}
    \end{subfigure}

    \caption{As illustrated in the figures, there is a clear self-clustering phenomenon within different sites, indicating that the model has learned shortcuts rather than features truly relevant to the disease. This is particularly evident on the zero-shot set: the MDD data from Site 1 significantly deviates from the distribution of the overall dataset, resulting in an inflated accuracy for the model.}
    \label{fig:de_mlp}
    
\end{figure}


\end{document}